\documentclass[twocolumn,showpacs,showkeys]{revtex4}
\usepackage{graphicx}
\usepackage{bm}
\usepackage{color}
\usepackage{amsmath}
\usepackage{natbib}

\begin{document}

\title{Dielectric permeability tensor and linear waves in spin-1/2 quantum kinetics with non-trivial equilibrium spin-distribution functions}

\author{Pavel A. Andreev}
\email{andreevpa@physics.msu.ru}
\author{L. S. Kuz'menkov}%
\email{lsk@phys.msu.ru}
\affiliation{Faculty of physics, Lomonosov Moscow State University, Moscow, Russian Federation.}

 \date{\today}

\begin{abstract}
A consideration of waves propagating parallel to the external magnetic field is presented. The dielectric permeability tensor is derived from quantum kinetic equations with non-trivial equilibrium spin-distribution functions in the linear approximation on amplitude of wave perturbations.
It is possible to consider equilibrium spin-distribution functions with nonzero z-projection proportional to the difference of the spin distribution function while x- and y-projections are equal to zero. It is called trivial equilibrium spin-distribution functions. In general case, x- and y-projections of the spin-distribution functions are nonzero which is called the non-trivial regime. Corresponding equilibrium solution is found in [Phys. Plasmas \textbf{23}, 062103 (2016)].
Contribution of the nontrivial part of the spin-distribution function appears in the dielectric permeability tensor in the additive form. It is explicitly found here.
Corresponding modification in the dispersion equation for the transverse waves is derived.
Contribution of nontrivial part of the spin-distribution function in the spectrum of transverse waves is calculated numerically.
It is found that the term caused by the nontrivial part of the spin-distribution
function can be comparable with the classic terms for the relatively small wave vectors and frequencies above the cyclotron frequency.
In majority of regimes, the extra spin caused term dominates over the spin term found earlier, except the small frequency regime, where their contributions in the whistler spectrum are comparable.
A decrease of the left-hand circularly polarized wave frequency, an increase of the high-frequency right-hand circularly polarized wave frequency, and
a decrease of frequency changing by an increase of frequency at the growth of the wave vector for the whistler are found.
A dramatic decrease of the spin wave frequency resulting in several times larger group velocity of the spin wave is found either.
Found dispersion equations are used for obtaining of an effective quantum hydrodynamics reproducing these results.
This generalization requires the introduction of corresponding equation of state for the thermal part of the spin current in the spin evolution equation.
\end{abstract}

%Use showkeys class option if keyword

\maketitle

%%%%%%%%%%TEXT

\section{Introduction}

Spin effects modify properties of plasmas \cite{Dyson PR 55, MaksimovTMP 2001, MaksimovTMP 2001 b, Brodin PRL 08 Cl Reg, Brodin PRL 08 g Kin, Andreev PoP kinetics 17 a, Andreev PoP kinetics 17 b, Dodin PRA 15 First-principle, Dodin PRA 15 Relativistic, Dodin PRA 17 Ponderomotive, Dodin PoP 17, Koide PRC 13, Ekman 1702, Mahajan IJTP 14} as well as electron gas in other mediums \cite{Barth JPC 72, Rajagopal PRB 73, Bloch ZP 29, Jones RMP 15}. They play a role for the degenerate and non-degenerate plasmas. However, the spin effects are more prominent for the degenerate plasmas since spin polarization clearly splits the Fermi step on two Fermi steps of different width for the spin-up and spin-down electrons \cite{Ryan PRB 91, Agarwal PRL 11, Andreev PRE 15 SEAW}. It allows to distinguish two types of electrons and consider them as two species \cite{Andreev PRE 15 SEAW, Kasuya PTP 56, Andreev AoP 15 SEAW}. Consideration of the independent evolution of electrons with different spin projections leads to discovery of the spin-electron acoustic waves.

Spin evolution modifies hydrodynamic and kinetic properties of plasma \cite{MaksimovTMP 2001, Lundin PRE 10, Andreev PRE 16, Hussain PP 14 spin bernst}. In both cases dynamical equations contain the force of spin-spin interaction: $S^{\beta}(\textbf{r},t)\nabla_{\textbf{r}}B^{\beta}$ in hydrodynamics \cite{Takabayasi comb}, $\nabla_{\textbf{p}}S^{\beta}(\textbf{r},\textbf{p},t)\cdot\nabla_{\textbf{r}}B^{\beta}$ in kinetics \cite{Dyson PR 55, Andreev kinetics 12, Andreev Phys A 15}, where $B^{\beta}=B^{\beta}(\textbf{r},t)$ is the magnetic field, $\nabla_{\textbf{r}}$ is the gradient in coordinate space, and $\nabla_{\textbf{p}}$ is the gradient in momentum space. This force contains the spin density $\textbf{S}$. It is the coordinate space density of spin in hydrodynamics $\textbf{S}(\textbf{r},t)$ and it is the phase space density of spin (the spin distribution function) in kinetics $\textbf{S}(\textbf{r},\textbf{p},t)$. Therefore, the complete model requires an equation of spin density evolution.
The time evolution of spin density happens due to two mechanisms: kinematic mechanism, where the flow of spinning particles in and out of the vicinity of the point in space change the local spin density, and dynamical, where the change of spin happens due to the interparticle interaction.
The kinematic mechanism gives the spin current.
In hydrodynamics, the spin current $J^{\alpha\beta}$ has the structure similar to the structure of the momentum current $\Pi^{\alpha\beta}$ existing in Euler equation \cite{MaksimovTMP 2001, Torrey PR 57, Andreev 1510 Spin Current}.
Tensor $\Pi^{\alpha\beta}$ contains the flow of the local center of mass $nv^{\alpha}v^{\beta}$, the flow on the thermal velocities (the thermal pressure or the Fermi pressure for degenerate fermions) $p^{\alpha\beta}$, and the quantum part which is usually called the quantum Bohm potential.
The spin current $J^{\alpha\beta}$ contains the flow of spin on the velocity of the local center of mass $S^{\alpha}v^{\beta}$, the thermal part of the spin current $J_{th}^{\alpha\beta}$ (or the Fermi spin current for degenerate fermions \cite{Andreev 1510 Spin Current}) and the quantum part calculated by Takabayasi \cite{Takabayasi comb}.

Majority of works on the spin evolution in plasmas are focused on interaction and drop the Fermi spin current \cite{Shukla UFN 10, Shukla RMP 11, Uzdensky RPP 14}.
The thermal spin current is not considered in the ferrofluids either \cite{Felderhof PRE 00}.
Hence, usually, the Fermi spin current is assumed to be equal to zero.
However, recently, an equation of state is derived for the Fermi spin current \cite{Andreev 1510 Spin Current}.
More detailed study of physical effects similar to the Fermi spin current required kinetic modeling.
Corresponding research is performed in Refs. \cite{Andreev PoP kinetics 17 a, Andreev PoP kinetics 17 b}.
It is shown that the kinetic analysis can be done with non-zero equilibrium scalar distribution function $f_{0}$ and non-zero z-projection of the equilibrium spin distribution function $S_{0z}$ while $S_{0x}=S_{0y}=0$ \cite{Andreev PoP kinetics 17 a, Andreev PoP kinetics 17 b}.
However, the general model requires consideration of $S_{0x}\neq0$  and $S_{0y}\neq0$ which are found in Ref. \cite{Andreev PoP 16 sep kin}.
Required analysis is performed in this paper for waves propagating parallel to the external magnetic field.

Influence of the spin on properties of magnetized plasmas is studied in many papers \cite{Dyson PR 55, MaksimovTMP 2001, MaksimovTMP 2001 b, Brodin PRL 08 Cl Reg, Marklund PRL07, Oraevsky AP 02, Oraevsky PAN}.
Quantum hydrodynamics \cite{MaksimovTMP 2001, MaksimovTMP 2001 b, Marklund PRL07, Andreev VestnMSU 2007} and quantum kinetics \cite{Dyson PR 55, Brodin PRL 08 Cl Reg, Lundin PRE 10, Andreev kinetics 12} are applied to this research.
There are considered ordinary electromagnetic waves \cite{Mahajan PoP 16}, spin damping corrections \cite{Asenjo PL A 09, Zhu PPCF 12}, waves propagating parallel \cite{Misra JPP 10},
and perpendicular \cite{Brodin PRL 08 g Kin, Andreev VestnMSU 2007, Andreev PoP 17 extr SEAWs, Andreev PoP 17 2D, Zamanian PoP 10} to the external magnetic field, and
oblique propagating waves \cite{Marklund PRL07, Asenjo PLA 12},
the quantum vorticity \cite{Yoshida JPA 16, Mahajan PRL 11, Andreev PP 15 Positrons, Braun PRL 12}, and the ponderomotive force \cite{Dodin PRA 15 Relativistic, Dodin PRA 17 Ponderomotive, Brodin PRL 10 SPF}.
Majority of these papers consider all electrons as one species.
Separate spin evolution quantum hydrodynamics and separate spin evolution quantum kinetics are developed for the study of electrons as two fluids \cite{Andreev PRE 15 SEAW, Andreev PoP 16 sep kin}.
The spin-electron acoustic waves found from these models are studied in different regimes \cite{Andreev AoP 15 SEAW, Andreev PRE 16, Andreev PoP 17 2D, Andreev EPL 16, Andreev APL 16, Andreev_Iqbal PoP 16}.
Study of the nontrivial part of the equilibrium distribution functions continues this research.

This paper is organized as follows.
In Sec. II, basic quantum kinetic equations for spin-1/2 plasmas are presented.
In Sec. III, the equilibrium distribution functions are presented and the main structure of the dielectric permeability tensor is described.
In Sec. IV, the dispersion equation and spectrum of the transverse waves propagating parallel to the external are studied under influence of extra spin effects caused by the non-trivial part of the equilibrium distribution functions.
In Sec. V, an equation of state for the Fermi spin current entering the hydrodynamic spin evolution equation is deducted from spectrum derived from kinetic model.
In Sec. VI, a summary of the obtained results is presented.
In Sec. VII, Appendix A is presented, where the linearized kinetic equations and their solutions are found.
In Sec. VIII, Appendix B is presented, where details of the the dielectric permeability tensor and some details of its calculation are described.
In Sec. IX, Appendix C is presented, where an approximate form of the dielectric permeability tensor is found.
In Sec. X, Appendix D is presented, where
the dimensionless form of the dispersion equation
is demonstrated.

\section{Quantum kinetic model for spin-1/2 plasmas}

The quantum kinetics of spin-1/2 particles can be modeled by the distribution functions (the scalar function $f$ and the vector (spin) function $\textbf{S}$) defined in the six-dimensional phase space \cite{Dyson PR 55, Andreev kinetics 12, Andreev Phys A 15, Hurst EPJD 14}.

Equation for the scalar distribution function $f=f(\textbf{r},\textbf{p},t)$ is the generalized Vlasov equation \cite{Dyson PR 55, Andreev kinetics 12, Andreev Phys A 15, Oraevsky AP 02, Oraevsky PAN, Hurst EPJD 14}:
$$\partial_{t}f+\textbf{v}\cdot\nabla_{\textbf{r}}f +q_{e}\biggl(\textbf{E}+\frac{1}{c}\textbf{v}\times\textbf{B}\biggr)\cdot\nabla_{\textbf{p}}f$$
\begin{equation}\label{SC_KA kinetic equation gen  classic limit with E and B}
+\mu_{e}\nabla^{\alpha}_{\textbf{r}} B^{\beta}\cdot\nabla_{\textbf{p}}^{\alpha} S^{\beta}=0,\end{equation}
which contains an extra term (the last term) caused by the spin-spin interaction.

The kinetic equation for the vector distribution function $\textbf{S}=\textbf{S}(\textbf{r},\textbf{p},t)$ has the following form \cite{Dyson PR 55, Andreev kinetics 12, Andreev Phys A 15, Oraevsky AP 02, Oraevsky PAN, Hurst EPJD 14}:
$$\partial_{t}S^{\alpha}+\textbf{v}\cdot\nabla_{\textbf{r}}S^{\alpha} +q_{e}\biggl(\textbf{E}+\frac{1}{c}\textbf{v}\times\textbf{B}\biggr)\cdot\nabla_{\textbf{p}}S^{\alpha}$$
\begin{equation}\label{SC_KA kinetic equation gen for spin classic limit with E and B}
+\mu_{e}\nabla^{\beta}_{\textbf{r}} B^{\alpha}\cdot\nabla_{\textbf{p}}^{\beta}f -\frac{2\mu_{e}}{\hbar}\varepsilon^{\alpha\beta\gamma}S^{\beta}B^{\gamma}=0.\end{equation}
The last two terms are caused by the spin-spin interaction. Kinetic equations (\ref{SC_KA kinetic equation gen  classic limit with E and B}) and (\ref{SC_KA kinetic equation gen for spin classic limit with E and B}) contain the following notations: $\textbf{E}$ and $\textbf{B}$ are the electric fields, $q_{e}=-\mid e\mid$ is the charge of electrons,
$\mu_{e}=-g\mu_{B}$ is the magnetic moment of electrons,
$\mu_{B}=\mid e\mid\hbar/2mc$ is the Bohr magneton, $g=1.00116$, $\textbf{r}$ ($\textbf{p}=m\textbf{v}$) is the coordinate in coordinate (momentum) space, $t$ is time, $\partial_{t}$ is the time derivative, $\nabla_{\textbf{r}}$ ($\nabla_{\textbf{p}}$) is the gradient on the space coordinate (on the momentum), $\nabla_{\textbf{p}}^{\alpha}$ and $\nabla_{\textbf{p}}^{\alpha}$ are projections of described gradients on the coordinate axis, $\hbar$ is the reduced Planck constant, $c$ is the speed of light, $\varepsilon^{\alpha\beta\gamma}$ is the antisymmetric symbol (the Levi–-Civita symbol).

Kinetic equations are coupled to the Maxwell equations
$\nabla\cdot \textbf{E}=4\pi\rho$, $\nabla\times \textbf{E}=-\frac{1}{c}\partial_{t}\textbf{B}$,  $\nabla\cdot \textbf{B}=0$, and
\begin{equation}\label{SC_KA rot B} \nabla\times \textbf{B}=\frac{1}{c}\partial_{t}\textbf{E}+\frac{4\pi}{c}\textbf{j} +4\pi\nabla\times \textbf{M},\end{equation}
where $\rho=q_{e}\int f(\textbf{r},\textbf{p},t)d\textbf{p}+q_{i}n_{0i}$, $\textbf{j}=q_{e}\int \textbf{v}f(\textbf{r},\textbf{p},t)d\textbf{p}$, $\textbf{M}=\mu_{e}\int \textbf{S}(\textbf{r},\textbf{p},t)d\textbf{p}$ is the magnetization.

\section{Structure of the dielectric permeability tensor}

Linear on the small perturbations kinetic equations are needed to be considered for the derivation of the dielectric permeability tensor. Assume that the following functions $f_{0}(p)$, $\textbf{S}_{0}(p,\varphi)$,
$\textbf{B}_{0}=\textbf{B}_{ext}=B_{0}\textbf{e}_{z}$ have non-zero values in the equilibrium state.
Moreover, the explicit forms of the equilibrium distribution functions for the partially spin polarized degenerate electrons appear as the sum or the difference of the Fermi steps for the spin-up and spin-down electrons \cite{Andreev PoP kinetics 17 a, Andreev PoP 16 sep kin}:
\begin{equation}\label{SC_KA equilib f} f_{0}(p)=\frac{1}{(2\pi\hbar)^{3}}[\Theta(p_{F\uparrow}-p) +\Theta(p_{F\downarrow}-p)], \end{equation}
and
\begin{equation}\label{SC_KA equilib x,y,z}  \begin{array}{ccc}
S_{0x}=\Sigma(p)\cos\varphi, & S_{0y}=\Sigma(p)\sin\varphi, & S_{0z}=\Sigma(p), \\
\end{array}\end{equation}
with
\begin{equation}\label{SC_KA equilib Sigma main} \Sigma(p)=\frac{1}{(2\pi\hbar)^{3}}[\Theta(p_{F\uparrow}-p) -\Theta(p_{F\downarrow}-p)],\end{equation}
where $\Theta$ is the step function, $p=|\textbf{p}|$ is the module of the momentum, $p_{F\uparrow}$ and $p_{F\downarrow}$ are the Fermi momentums for the spin-up and spin-down electrons, $v_{Fs}=p_{Fs}/m=(6\pi^{2}n_{0s})^{1/3}\hbar/m$, $s=\uparrow, \downarrow$, and $\varphi$ is the polar angle in cylindrical coordinates in the momentum space, $n_{0s}$ are the concentrations of the spin-up and spin-down electrons.
Present small perturbations of the described equilibrium state as the plane waves. For instance, the scalar distribution function can be presented as $f=f_{0}+\delta f$, where $\delta f=F e^{-\imath\omega t+\imath \textbf{k}\textbf{r}}$ and $F$ is an amplitude of perturbation.
Moreover, the waves propagating parallel to the external magnetic field are under consideration $\textbf{k}=\{0,0,k_{z}\}$.
Below, the following notations are used for the charge cyclotron frequency $\Omega_{e}=q_{e}B_{0}/mc$, the magnetic moment cyclotron frequency $\Omega_{\mu}=2\mu_{e}B_{0}/\hbar$, and the partial Langmuir frequency $\omega_{Ls}^{2}=4\pi e^{2}n_{0s}/m$.
Parameters $\Omega_{e}=q_{e}B_{0}/mc$ and $\Omega_{\mu}=2\mu_{e}B_{0}/\hbar$ are equal to each other if the anomalous part of magnetic moment of electron is neglected.
Linearized kinetic equations and their solutions are presented in Appendix A.

Linear kinetic theory allows to calculate the dielectric permeability tensor which enters the equation for the small perturbations of the electric field:
\begin{equation}\label{SC_KA } \biggl[k^{2}\delta^{\alpha\beta}-k^{\alpha}k^{\beta}-\frac{\omega^{2}}{c^{2}}\varepsilon^{\alpha\beta}(\omega)\biggr]\delta E_{\beta}=0, \end{equation}
where the dielectric permeability tensor appears as
\begin{equation}\label{SC_KA }\varepsilon^{\alpha\beta}(\omega)=\delta^{\alpha\beta}+\varepsilon^{\alpha\beta}_{1}(\omega)+\varepsilon^{\alpha\beta}_{2}(\omega),\end{equation}
where $\delta^{\alpha\beta}$ is the Kronecker symbol, $\varepsilon^{\alpha\beta}_{1}(\omega)$ is
the dielectric permeability tensor caused by the current
\begin{equation}\label{SC_KA sigma 1 short}\varepsilon^{\alpha\beta}_{1}(\omega)\delta E_{\beta}=\frac{4\pi\imath}{\omega}\frac{q_{e}}{m}\int p^{\alpha}\delta f d\textbf{p},\end{equation}
and $\varepsilon^{\alpha\beta}_{2}(\omega)$ is the dielectric permeability tensor caused by the curl of magnetization
\begin{equation}\label{SC_KA sigma 2 short}\varepsilon^{\alpha\beta}_{2}(\omega)\delta E_{\beta}=\frac{4\pi\mu_{e}}{\omega}\varepsilon^{\alpha\gamma z} k_{z}c \int\delta S^{\gamma}d\textbf{p}.\end{equation}

The explicit form of $\varepsilon^{\alpha\beta}_{1}$ and
$\varepsilon^{\alpha\beta}_{2}$ are presented in Appendix B.
Meanwhile, consider structure of tensors $\varepsilon^{\alpha\beta}_{1}$ and
$\varepsilon^{\alpha\beta}_{2}$.
Tensor $\varepsilon^{\alpha\beta}_{1}$ can be separated into three parts: $\varepsilon^{\alpha\beta}_{1}=\varepsilon^{\alpha\beta}_{10}+\varepsilon^{\alpha\beta}_{11}+\varepsilon^{\alpha\beta}_{12}$.
The first part is the quasi-classic term $\varepsilon^{\alpha\beta}_{10}$ existing with no account of the spin evolution.
Tensor $\varepsilon^{\alpha\beta}_{10}$ consists of two terms $\varepsilon^{\alpha\beta}_{10}=-\sum_{s=\uparrow, \downarrow}\int \sin\theta d\theta \widetilde{\Pi}^{\alpha\beta}_{Cl}(\theta,s)$ since the spin polarized plasma with the equilibrium scalar distribution function splitted in two terms (\ref{SC_KA equilib f}) is considered, where $\theta$
is an angle of the spherical coordinates in the
velocity space defined as $\cos\theta=v_{z}/v$.
Despite this fact tensor $\varepsilon^{\alpha\beta}_{10}$ has well-known structure presented in textbooks (see for instance \cite{Rukhadze book 84}).
Tensor $\varepsilon^{\alpha\beta}_{2}$ can be separated into two parts: $\varepsilon^{\alpha\beta}_{2}=\varepsilon^{\alpha\beta}_{21}+\varepsilon^{\alpha\beta}_{22}$.
Tensors
$\varepsilon^{\alpha\beta}_{11}$, $\varepsilon^{\alpha\beta}_{12}$, $\varepsilon^{\alpha\beta}_{21}$, $\varepsilon^{\alpha\beta}_{22}$ appear due to the spin evolution.
Tensors $\varepsilon^{\alpha\beta}_{11}$ and $\varepsilon^{\alpha\beta}_{21}$ are found at the account of the trivial part of the equilibrium distribution functions.
It means that $f_{0}$ and $S_{0z}$ are given by
equations (\ref{SC_KA equilib f}) and
(\ref{SC_KA equilib x,y,z}) while $S_{0x}=S_{0y}=0$.
The account of non-zero $S_{0x}$, $S_{0y}$ given by
equations (\ref{SC_KA equilib x,y,z})
(the nontrivial part of the equilibrium distribution functions) leads to
existence of tensors $\varepsilon^{\alpha\beta}_{12}$, $\varepsilon^{\alpha\beta}_{22}$. Calculation of tensors $\varepsilon^{\alpha\beta}_{12}$ and $\varepsilon^{\alpha\beta}_{22}$ and derivation of their contribution to the plasma properties are the main subjects of this paper.

The parts of the dielectric permeability tensor by
$\varepsilon^{\alpha\beta}_{10}$, $\varepsilon^{\alpha\beta}_{11}$, $\varepsilon^{\alpha\beta}_{21}$
are in accordance with the earlier developed models
\cite{Andreev PoP kinetics 17 a, Andreev PoP kinetics 17 b, Lundin PRE 10, Hurst EPJD 14},
while tensors
$\varepsilon^{\alpha\beta}_{12}$ and $\varepsilon^{\alpha\beta}_{22}$ are the generalization of the mentioned papers.

\section{Transverse waves propagating parallel to the external magnetic field}

A partially explicit form of dispersion equation for the
transverse waves appears as follows  %from equations (\ref{SC_KA tr waves general form})
$$\frac{k_{z}^{2}c^{2}}{\omega^{2}}=1 -\sum_{s=\uparrow, \downarrow}
\frac{3}{4}\frac{\omega_{Ls}^{2}}{\omega} \frac{1}{k_{z}v_{Fs}}\biggl[\frac{2(\omega\mp\Omega_{e})}{k_{z}v_{Fs}}$$
\begin{equation}\label{SC_KA tr waves explicit form} +\biggl(1-\frac{(\omega\mp\Omega_{e})^{2}}{(k_{z}v_{Fs})^{2}}\biggr) \ln\biggl(\frac{\omega+k_{z}v_{Fs}\mp\Omega_{e}}{\omega-k_{z}v_{Fs}\mp\Omega_{e}}\biggr)\biggr] +\Sigma_{\mp}, \end{equation}
for $\delta E_{x}=\pm\imath\delta E_{y}$ (left/right hand circular polarization) correspondingly,
where $\Sigma_{\mp}$ are the terms caused by the spin evolution presented in the nonexplicit form.

\begin{figure}
\includegraphics[width=8cm,angle=0]{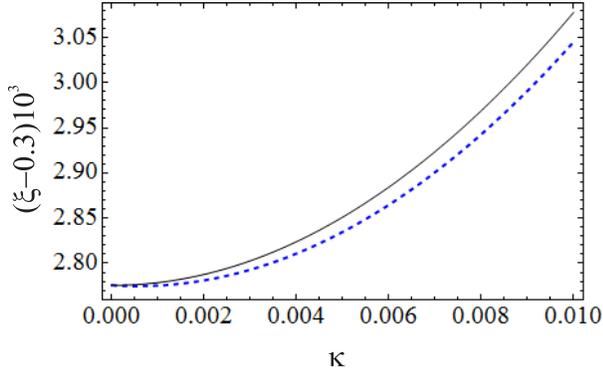}
\caption{\label{susdSOI 01} (Color online) The figure shows spectrum of transverse wave in plasma with the left-hand circular polarization $\delta E_{x}=\imath\delta E_{y}$ at $\mid\Omega_{e}\mid/\omega_{Le}=3$, $\eta=0.9$, and $n_{0}=10^{27}$ cm$^{-3}$. The following dimensionless notations are used in this figure and all figures below: $\xi=\omega/\omega_{Le}$, $\kappa=k_{z}c/\omega_{Le}$.
The continuous (black) line shows the spectrum without the spin account.
The dashed (blue) 
line shows the spectrum with the account of spin evolution.
It appears as a solution of equation (\ref{SC_KA tr waves explicit form expanded  reduced 2}) with the upper sign.}
\end{figure}
\begin{figure}
\includegraphics[width=8cm,angle=0]{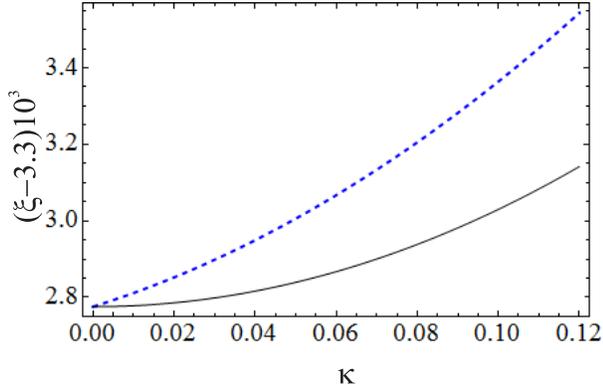}
\caption{\label{susdSOI 02} (Color online) The figure shows spectrum of (upper) classic wave with the right-hand circular polarization $\delta E_{x}=-\imath\delta E_{y}$ at $\mid\Omega_{e}\mid/\omega_{Le}=3$, $\eta=0.9$, and $n_{0}=10^{27}$ cm$^{-3}$.
The continuous (black) line shows the spectrum without the spin account.
The dashed (blue)
line shows the spectrum with the account of spin evolution.
It appears as a solution of equation (\ref{SC_KA tr waves explicit form expanded  reduced 2}) with the lower sign.}
\end{figure}
\begin{figure}
\includegraphics[width=8cm,angle=0]{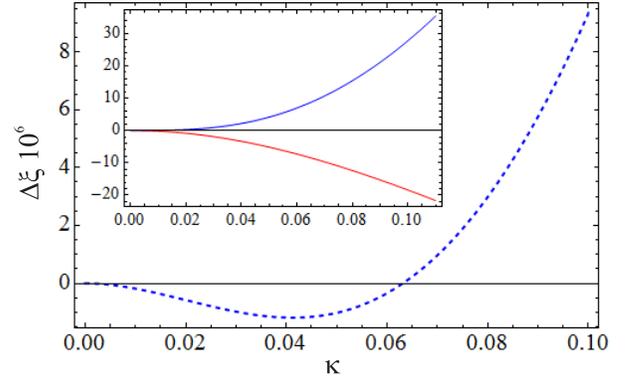}
\caption{\label{susdSOI 03} (Color online) The figure shows the spectrum shift of (lower) transverse classic wave (whistler) with the right-hand circular polarization $\delta E_{x}=-\imath\delta E_{y}$ at $\mid\Omega_{e}\mid/\omega_{Le}=3$, $\eta=0.9$, and $n_{0}=10^{27}$ cm$^{-3}$.
It appears as a solution of the equation (\ref{SC_KA tr waves explicit form expanded  reduced 2 minus}).
The dashed line shows the spectrum shift under the influence of spin evolution.
The small figure shows shifts of frequency caused by each of spin terms independently.
The third (the last) term on the right-hand side of equation (\ref{SC_KA tr waves explicit form expanded  reduced 2 minus}) gives negative (positive) contribution.}
\end{figure}

\begin{figure}
\includegraphics[width=8cm,angle=0]{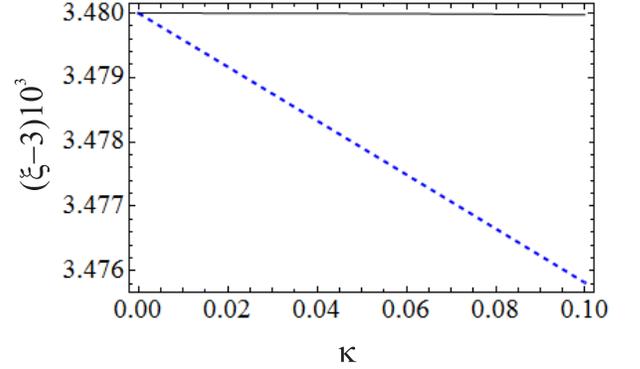}
\caption{\label{susdSOI 05} (Color online) The figure shows spectrum of the spin wave at $\mid\Omega_{e}\mid/\omega_{Le}=3$, $\eta=0.9$, and $n_{0}=10^{27}$ cm$^{-3}$, where  $\delta E_{x}=-\imath\delta E_{y}$ (the right-hand circular polarization).
The continuous line shows the spin wave spectrum found from the trivial distribution functions.
The dashed line is found at the account of the non-trivial part of the distribution function.
It appears as a solution of equation (\ref{SC_KA tr waves explicit form expanded  reduced 2}) with the lower sign.}
\end{figure}

Consider new term in the long-wavelength regime $k_{z}v_{Fs}/\mid\omega\pm\Omega_{e}\mid\ll1$. Start with regime $\omega\pm\mid\Omega_{e}\mid>0$. In this regime, equation (\ref{SC_KA tr waves explicit form}) has the following form
$$\frac{k_{z}^{2}c^{2}}{\omega^{2}}=1
-\frac{\omega_{Le}^{2}}{\omega (\omega\pm\mid\Omega_{e}\mid)}
+\frac{\omega_{Le}^{2}}{\omega^{2}}\frac{\hbar k_{z}^{2}}{2m n_{0e}} \biggl[\frac{n_{0\uparrow}-n_{0\downarrow}}{\omega\pm\mid\Omega_{\mu}\mid}\biggr]$$
\begin{equation}\label{SC_KA tr waves explicit form expanded  reduced 2}
\mp\frac{ (6\pi^{2})^{\frac{2}{3}}}{32\pi} \frac{\omega_{Le}^{2}}{\omega(\omega\pm\mid\Omega_{\mu}\mid)}\frac{n_{0\uparrow}^{\frac{2}{3}} -n_{0\downarrow}^{\frac{2}{3}}}{n_{0e}}k_{z} ,\end{equation}
where $\omega_{Le}^{2}=\omega_{L\uparrow}^{2}+\omega_{L\downarrow}^{2}=4\pi e^{2}n_{0e}/m$ is the Langmuir frequency for all electrons and $n_{0e}=n_{0\uparrow}+n_{0\downarrow}$ is the concentration of all electrons.

Equation (\ref{SC_KA tr waves explicit form expanded  reduced 2}) contains coefficients proportional to $\hbar k_{z}^{2}/m$. It resembles a similarity to the well-known quantum Bohm potential. However, it comes from the spin evolution \cite{Misra JPP 10}.
Equation (\ref{SC_KA tr waves explicit form expanded  reduced 2})
is a generalization of the equation which is well-known from spin-1/2 hydrodynamics \cite{Andreev VestnMSU 2007, Misra JPP 10}.

The fourth term on the right-hand side of equation (\ref{SC_KA tr waves explicit form expanded  reduced 2}) is proportional to the difference of the Fermi energies for electrons with different spin projections $\varepsilon_{F\uparrow}-\varepsilon_{F\downarrow}\sim n_{0\uparrow}^{\frac{2}{3}} -n_{0\downarrow}^{\frac{2}{3}}$ which is a signature of the Fermi spin current (see equation of state derived in \cite{Andreev 1510 Spin Current}, and discussion in introductions of Refs. \cite{Andreev PoP kinetics 17 a, Andreev PoP kinetics 17 b}).

Compare the fourth term on the right-hand side of equation (\ref{SC_KA tr waves explicit form expanded  reduced 2}) with $k_{z}^{2}c^{2}/\omega^{2}$.
Consider the ration of these terms and find
$\Xi=(3\pi^{2})^{\frac{2}{3}} r\tau (n_{0e}^{\frac{1}{3}}/k_{z})\omega/8(\omega\pm\mid\Omega_{\mu}\mid)$,
where $r=e^2 n_{0e}^{\frac{1}{3}}/mc^2$ and
$\tau=[(1+\eta)^{\frac{2}{3}}-(1-\eta)^{\frac{2}{3}}]$
with $\eta=\mid n_{0\uparrow}-n_{0\downarrow}\mid/n_{0e}\in[0,1]$ is the spin polarization.
Basically, this ratio is proportional to parameter $r$.
Parameter $r$ is small even for $n_{0}=10^{27}$ cm$^{-3}$ ($r\approx2\times10^{-4}$).
The ratio $\Xi$ is not affected by frequency in the high-frequency regime $\omega\gg\mid\Omega_{\mu}\mid$.
It is decreased by $\omega/\mid\Omega_{\mu}\mid$ at small frequencies while $\Xi$ grows at the intermediate frequencies $\omega\approx\mid\Omega_{\mu}\mid$ for the right-hand circular polarization.
Small spin polarization $\eta\ll1$ decreases parameter $\Xi$.
The ratio $\Xi$ increases at the large spin polarization $\eta\sim1$ and small wave vectors $k_{z}\ll\sqrt[3]{n_{0e}}$.

Next, compare the second term and the fourth term on the right-hand side of equation (\ref{SC_KA tr waves explicit form expanded  reduced 2}). Their ratio has the following form $\Lambda=(3\pi^{2})^{\frac{2}{3}}(k_{z}/n_{0e}^{\frac{1}{3}})\tau/32\pi$. %{\color{blue}}
Ratio $\Lambda$ grows with increase of wave vector $k_{z}$, but large values of $k_{z}$ cannot be considered since equation (\ref{SC_KA tr waves explicit form expanded  reduced 2}) is derived in the following limit $\omega\pm\mid\Omega_{a}\mid\gg k_{z}v_{Fs}$. Find $\Lambda\sim10^{-2}$ at $k_{z}=0.1n_{0}^{\frac{1}{3}}$, $\eta=0.5$, and $\omega\gg\mid\Omega_{a}\mid$. It is relatively small value, but it is a good value for a spin effect in plasma. In this regime $\Lambda\gg\Xi$.

Moreover,
compare the third term and the fourth term on the right-hand side
of equation (\ref{SC_KA tr waves explicit form expanded  reduced 2}).
Both of them present contribution of spin effects while the fourth term is
derived in this paper. %{\color{green}}
Their ratio has the following form
\begin{equation}\label{SC_KA} \Pi=\frac{8\pi}{(3\pi^{2})^{\frac{2}{3}}}\frac{k_{z}}{n_{0e}^{\frac{1}{3}}} \frac{\hbar n_{0e}^{\frac{2}{3}}}{m\omega}\frac{(1+\eta)^{\frac{2}{3}}+(1-\eta^{2})^{\frac{1}{3}}+(1-\eta)^{\frac{2}{3}}}{(1+\eta)^{\frac{1}{3}}+(1-\eta)^{\frac{1}{3}}}\end{equation}
It is decreased by factor $k_{z}/n_{0e}^{1/3}$, but for small frequencies $\omega$ parameter $\hbar n_{0e}^{2/3}/(m\omega)$ leads to increase of $\Pi$.

Consider the left-hand circularly polarized electromagnetic waves taking the upper sign in equation (\ref{SC_KA tr waves explicit form expanded  reduced 2}). This wave appears due to the classic terms in equation (\ref{SC_KA tr waves explicit form expanded  reduced 2}). Consider modifications of its properties arising due to the spin terms: the third and fourth terms on the right-hand side of equation (\ref{SC_KA tr waves explicit form expanded  reduced 2}).
The third term ($\sim\hbar k_{z}^{2}$) is small in considering regime of the small wave vectors.
Hence, this analysis is focused on the last term in equation (\ref{SC_KA tr waves explicit form expanded  reduced 2}).
It has positive sign ($n_{0\uparrow}<n_{0\downarrow}$) while the classic term coming from the charge evolution (the second term on the right-hand side) is negative.
Hence, they give opposite influences.
The classic term gives a considerable increase of frequency in compare with the frequency of wave propagating in vacuum.
So, the spin term decreases the frequency of the wave.
This effect is presented in Fig. \ref{susdSOI 01}.
The spin caused third and fourth terms have opposite signs in this regime.

Next, consider the right-hand circularly polarized electromagnetic waves.
In this regime, the lower sign should be taken in equation (\ref{SC_KA tr waves explicit form expanded  reduced 2}) (for $\omega>\mid\Omega_{e}\mid$).
Consider frequencies $\omega$ larger $1.1\mid\Omega_{e}\mid$ and obtain the classic right-hand circularly polarized electromagnetic wave.
Its analysis is similar to the presented above for the left-hand circularly
polarized waves, but the sign of the last term in equation (\ref{SC_KA tr waves explicit form expanded  reduced 2}) is different.
It is negative like the classic term caused by the electron motion.
Therefore, both terms lead to the increase of the frequency.
The increase of the frequency of right-hand circularly polarized electromagnetic wave caused by the spin effects in compare to the classic regime is demonstrated in Fig. \ref{susdSOI 02}.
The spin caused third and fourth terms have same sign in this regime.

The regime of right-hand circularly polarized waves demonstrates the spin wave solution at $0<\omega-\mid\Omega_{e}\mid<0.1\mid\Omega_{e}\mid$.

Equation (\ref{SC_KA tr waves explicit form expanded  reduced 2}) corresponds to relatively large deviation of frequency $\omega$ from the cyclotron frequency $\mid\Omega_{e}\mid$ for the large variations of the wave vector, or the small deviations of frequency $\omega$ from the cyclotron frequency $\mid\Omega_{e}\mid$ for the small values of the wave vector.
Hence, it allows consideration of area $\omega\approx\mid\Omega_{e}\mid$ and analysis of the spin waves with $\omega\approx\mid\Omega_{\mu}\mid$ for small wave vectors only. It is illustrated in Fig. \ref{susdSOI 05}.
It shows that the nontrivial part of the equilibrium distribution functions increases the deviation of the spin wave frequency from the cyclotron frequency several times in compare with the results following from the third term on the right-hand side in equation (\ref{SC_KA tr waves explicit form expanded  reduced 2}).

Next, consider regime $\omega-\mid\Omega_{e}\mid<0$, which is meaningful for the right-hand circularly polarized waves.
As the result find, find the following dispersion equation:
$$\frac{k_{z}^{2}c^{2}}{\omega^{2}}=1
-\frac{\omega_{Le}^{2}}{\omega (\omega-\mid\Omega_{e}\mid)}
+\frac{\omega_{Le}^{2}}{\omega^{2}}\frac{\hbar k_{z}^{2}}{2m n_{0e}} \biggl[\frac{n_{0\uparrow}-n_{0\downarrow}}{\omega-\mid\Omega_{\mu}\mid}\biggr]$$
\begin{equation}\label{SC_KA tr waves explicit form expanded  reduced 2 minus}
-\frac{ (6\pi^{2})^{\frac{2}{3}}}{32\pi} \frac{\omega_{Le}^{2}}{\omega(\omega-\mid\Omega_{\mu}\mid)}\frac{n_{0\uparrow}^{\frac{2}{3}} -n_{0\downarrow}^{\frac{2}{3}}}{n_{0e}}k_{z} .\end{equation}

Equation (\ref{SC_KA tr waves explicit form expanded  reduced 2 minus}) demonstrates that the last term changes its sign for the right-hand circular polarized waves at the transition to the small frequency regime $\omega<\mid\Omega_{\mu}\mid$
in compare with the large frequency regime $\omega>\mid\Omega_{\mu}\mid$ presented by equation
(\ref{SC_KA tr waves explicit form expanded  reduced 2}) with the lower sign.

Considering the low-frequency limit of the dispersion equation for the
right-hand circularly polarized waves (whistlers)
(\ref{SC_KA tr waves explicit form expanded  reduced 2 minus}),
find the following analytical solution:
$$\omega=\mid\Omega_{e}\mid
\Biggl[ \frac{k_{z}^{2}c^{2}}{\omega^{2}_{Le}}-\eta\frac{\hbar k_{z}^{2}}{2m\mid\Omega_{e}\mid}
$$
\begin{equation}\label{SC_KA whistler} +
\frac{k_{z}^{2}c^{2}}{\omega^{2}_{Le}}
\Biggl(\frac{(3\pi^{2})^{\frac{2}{3}}}{32\pi} \biggl((1+\eta)^{\frac{2}{3}}-(1-\eta)^{\frac{2}{3}}\biggr)
\frac{k_{z}}{n_{0e}^{\frac{1}{3}}}\Biggr)
\Biggr]. \end{equation}
It is found by the iteration method assuming that spin contribution is relatively small.
The first term in equation (\ref{SC_KA whistler}) is the classic term.
The spin effects give two contributions in equation (\ref{SC_KA whistler}).
The second term on the right-hand side is found earlier in literature \cite{Andreev VestnMSU 2007, Misra JPP 10}.
In this model it comes from the trivial part of the equilibrium distribution functions.
Its contribution gives a decrease of frequency of the whistler.
The last term in equation (\ref{SC_KA whistler}) comes from the non-trivial part of the equilibrium distribution functions and gives an increase of the whistler frequency
as it is demonstrated in Fig. \ref{susdSOI 03}.

As it is described above, at small frequencies the last term is decreased by factor $\omega/\mid\Omega_{e}\mid$. Hence, it becomes comparable with the third term and competition between them is revealed in nonmonotonial shift of the whistler spectrum (see Fig. \ref{susdSOI 03}).

\section{Hydrodynamic Fermi spin current}

Analysis shows that a phenomenological generalization of the quantum hydrodynamic equations allows to derive the last term in equations (\ref{SC_KA tr waves explicit form expanded  reduced 2}), (\ref{SC_KA tr waves explicit form expanded  reduced 2 minus}), and (\ref{SC_KA whistler}) which is the main spin correction appearing in the considered regime. It is found that an additional term should appear in the spin (magnetization) evolution equation. Moreover, it can be interpreted as the divergence of the spin current (spin flux).
Therefore, an equation of state for the hydrodynamic Fermi spin current existing in the hydrodynamic spin evolution equation can be extracted from obtained results.
The equation of state corresponds to the long-wavelength $k\rightarrow0$ limits.
In this limit, the Fermi spin current leading to the last terms in equations (\ref{SC_KA tr waves explicit form expanded  reduced 2}) and  (\ref{SC_KA tr waves explicit form expanded  reduced 2 minus}) can be captured for future study of the long-wavelength excitations.
The magnetization evolution equation has the following structure \cite{MaksimovTMP 2001, Andreev 1510 Spin Current, Mahajan PRL 11}
$$n(\partial_{t}+\textbf{u}\cdot\nabla) \mbox{\boldmath $\mu$}$$
\begin{equation}\label{SC_KA }  -\frac{\hbar}{2m\mu_{e}}\partial^{\beta}[n\mbox{\boldmath $\mu$}\times \partial^{\beta}\mbox{\boldmath $\mu$} ] +\mbox{\boldmath $\Im$}=\frac{2\mu_{e}}{\hbar}n[\mbox{\boldmath $\mu$}\times\textbf{B}], \end{equation}
where $\mbox{\boldmath $\mu$}=\textbf{M}/n$, $\textbf{M}=\textbf{M}(\textbf{r},t)$ is the magnetization, $n=n(\textbf{r},t)$ is the concentration of particles, $\textbf{u}(\textbf{r},t)$ is the velocity field, and $\mbox{\boldmath $\Im$}$ is the divergence of the thermal part of the spin current (which is called Fermi spin current for the degenerate electron gas), its explicit form is found in the following form
\begin{equation}\label{SC_KA spin current x} \Im_{x}= \mp\frac{(6\pi^{2})^{\frac{2}{3}}}{32\pi}
\frac{\omega_{Le}^{2}}{(\omega\pm\mid\Omega_{\mu}\mid) c}\frac{n_{0\uparrow}^{\frac{2}{3}}-n_{0\downarrow}^{\frac{2}{3}}}{n_{0}^{2}} (\Omega_{\mu}\delta E_{x}+\imath\omega\delta E_{y}) ,\end{equation}
\begin{equation}\label{SC_KA spin current x} \Im_{y}= \mp\frac{(6\pi^{2})^{\frac{2}{3}}}{32\pi}
\frac{\omega_{Le}^{2}}{(\omega\pm\mid\Omega_{\mu}\mid)c}
\frac{ n_{0\uparrow}^{\frac{2}{3}}-n_{0\downarrow}^{\frac{2}{3}} }{n_{0}^{2}} (\imath\omega\delta E_{x}-\Omega_{\mu}\delta E_{y}) ,\end{equation}
$\Im_{z}=0$, for left/right-hand polarized transverse waves correspondingly,
at $\omega\pm\mid\Omega_{\mu}\mid>0$.
At $\omega<\mid\Omega_{\mu}\mid$,
the lower sing should be chosen in the denominator and the upper sing should be chosen in front of the expression.
This is a frequency dependent equation of state. Hence, it can be suitable for the linear or weakly-non-linear phenomena.

Advantage of this equation of state is the fact that it derived for the perturbation evolution while other equations of state derived in literature are derived for the equilibrium regimes
\cite{Andreev 1510 Spin Current, Andreev PoP 16 sep kin}. However, as it is mentioned above, the equation of state is found in small range of parameters.

\section{Conclusion}

It has been demonstrated that the non-trivial part of the equilibrium distribution functions gives a considerable contribution in the long-wavelength limit. Spectra of all classic transverse waves propagating parallel to the external field are changed.
In this regime, there are three classic waves and the spin wave.
The spectra of classic waves are shifted
while spin wave spectrum is modified dramatically.
The well-known dispersion equation follows from the trivial part of the distribution functions leads to decreasing frequency as a function of the wave vector.
The non-trivial part of the equilibrium distribution functions leads to further decrease of frequency as a function of the wave vector.
Moreover, the module of group velocity $d\omega/dk_{z}$ increases several times.

It has been found that the dispersion equation is different for the right-hand circularly polarized waves with $\omega<\mid\Omega_{e}\mid$
while the hydrodynamic model or the earlier developed kinetic models give the same dispersion equation in both regimes.
The modifications are caused by the by the non-trivial part of the equilibrium distribution functions.
Corresponding generalization of hydrodynamic equations has been developed,
where the frequency dependent equation of state for the spin current has been found and included.

An analytical expression is found for the spectrum of whistler. This spectrum explicitly presents the contribution of the spin effects including effects caused by the non-trivial equilibrium distribution functions.

All described above show that the developed kinetic model is necessary for the description of the spin effects in plasmas.
This model is an essential generalization of existing models.
The model allows to discover new spin related effects in linear and
non-linear waves propagating parallel or perpendicular to the external magnetic field or for the oblique wave propagation.
The equation of state for the hydrodynamic spin current has been extracted from the obtained results.
Hence, the generalized hydrodynamic will provide a simple approximate description of phenomena related to the nontrivial part of the equilibrium distribution functions.

\begin{acknowledgements}
The work of P.A. was supported by the Russian
Foundation for Basic Research (grant no. 16-32-00886) and the Dynasty foundation.
\end{acknowledgements}

\section{Appendix A: Linear solutions of kinetic equations}

The following linearized Fourier transformed kinetic equations can be found from kinetic equations (\ref{SC_KA kinetic equation gen  classic limit with E and B}) and (\ref{SC_KA kinetic equation gen for spin classic limit with E and B}) at the consideration of small amplitude plane wave perturbations of the described equilibrium state:
$$-\imath\omega\delta f
+\imath \textbf{v}\cdot\textbf{k}\delta f +\frac{q_{e}}{c}B_{0} (\textbf{v}\times \textbf{e}_{z})\cdot\nabla_{\textbf{p}}\delta f$$
\begin{equation}\label{SC_KA kin eq f lin 3D} +q_{e}\delta \textbf{E}\cdot\nabla_{\textbf{p}}f_{0}+\imath\mu_{e}(\textbf{k}\cdot\nabla_{\textbf{p}})(\textbf{S}_{0}\cdot\delta \textbf{B})=0,\end{equation}
and
$$-\imath\omega\delta \textbf{S}+\imath (\textbf{v}\cdot\textbf{k}) \delta \textbf{S}
+\frac{q_{e}}{c}B_{0} ((\textbf{v}\times \textbf{e}_{z})\cdot\nabla_{\textbf{p}})\delta \textbf{S}$$
$$+\frac{q_{e}}{c}((\textbf{v}\times\delta \textbf{B})\cdot\nabla_{\textbf{p}})\textbf{S}_{0}+\imath\mu_{e} (\textbf{k}\cdot\nabla_{\textbf{p}})f_{0}\delta \textbf{B}$$
\begin{equation}\label{SC_KA kin eq S lin 3D} +q_{e}(\delta \textbf{E}\cdot\nabla_{\textbf{p}})\textbf{S}_{0} +\frac{2\mu_{e}}{\hbar}(\textbf{B}_{0}\times\delta \textbf{S}-\textbf{S}_{0}\times \delta \textbf{B})=0,\end{equation}
where the wave vector has the following structure $\textbf{k}=\{0,0,k_{z}\}$ that corresponds to the propagation of waves parallel to the external magnetic field which leads to $\delta B_{z}=0$. Consider the following part of the Lorentz-like force in the spin evolution kinetic equation $\frac{q_{e}}{c}((\textbf{v}\times\delta \textbf{B})\cdot\nabla_{\textbf{p}})\textbf{S}_{0}$ which is non-zero since $S_{0x}$ and $S_{0y}$ are non-isotropic functions. It can be represented in the following form $-(q_{e}/mc)\varepsilon^{\alpha z\gamma}S_{0\gamma}(\textbf{v}^{2}\delta B_{z}-v_{z}(\textbf{v}\cdot\delta \textbf{B})/v_{\perp}^{2}$.
Present more explicit form of the sixth term in the spin evolution kinetic equation $q_{e}(\delta \textbf{E}\cdot\nabla_{\textbf{p}})\textbf{S}_{0}=q_{e}(\delta \textbf{E}\cdot\textbf{v})\partial_{\varepsilon}\textbf{S}_{0}+(\textbf{e}_{z}\times \textbf{S}_{0}) ((\textbf{v}\times\delta \textbf{E})\cdot\textbf{e}_{z})/mv_{\perp}^{2}$,
where $\partial_{\varepsilon}$ is the derivative on kinetic energy $\varepsilon=\textbf{p}^{2}/2m$.

Solution of the linearized kinetic equations (\ref{SC_KA kin eq f lin 3D}) and (\ref{SC_KA kin eq S lin 3D}) leads to the following perturbations of the distribution functions
\begin{widetext}
\begin{equation}\label{SC_KA f solution ext}\delta f=\frac{1}{\Omega_{e}}\int_{C_{0}}^{\varphi}\biggl(q_{e}(\textbf{v}\cdot\delta \textbf{E})\frac{\partial f_{0}}{\partial\varepsilon} +\imath\mu_{e}(k_{z}v_{z})\biggl(\delta \textbf{B}\cdot \frac{\partial \textbf{S}_{0}}{\partial\varepsilon}\biggr)\biggr)
\exp\biggl(\imath\frac{(\omega-k_{z}v_{z})}{\Omega_{e}}(\varphi'-\varphi)\biggr)d\varphi',\end{equation}

\begin{equation}\label{SC_KA S z solution}
\delta S_{z}=\frac{1}{\Omega_{e}}\int_{C_{3}}^{\varphi}\biggl(q_{e}(\textbf{v}\cdot\delta \textbf{E})\frac{\partial S_{0z}}{\partial\varepsilon}
+\frac{2\mu_{e}}{\hbar}(\delta B_{x}S_{0y}-\delta B_{y}S_{0x})\biggr)\exp\biggl(\imath\frac{(\omega-k_{z}v_{z})}{\Omega_{e}}(\varphi'-\varphi)\biggr)d\varphi',\end{equation}

$$\delta S_{x}=\frac{1}{2}\Biggl[ \int_{C_{1}}^{\varphi}\exp\biggl(-\frac{\imath\Omega_{\mu}}{\Omega_{e}}(\varphi'-\varphi)\biggr)(\Pi_{x}(\varphi')+\imath\Pi_{y}(\varphi'))d\varphi'
$$
\begin{equation}\label{SC_KA anzac Sx} +\int_{C_{2}}^{\varphi}\exp\biggl(\frac{\imath\Omega_{\mu}}{\Omega_{e}}(\varphi'-\varphi)\biggr)(\Pi_{x}(\varphi')-\imath\Pi_{y}(\varphi'))d\varphi'\Biggr] \exp\biggl(-\imath\frac{(\omega-k_{z}v_{z})}{\Omega_{e}}\varphi\biggr), \end{equation}
and
$$\delta S_{y}=\frac{1}{2\imath}\Biggl[ \int_{C_{1}}^{\varphi}\exp\biggl(-\frac{\imath\Omega_{\mu}}{\Omega_{e}}(\varphi'-\varphi)\biggr)(\Pi_{x}(\varphi')+\imath\Pi_{y}(\varphi'))d\varphi'
$$
\begin{equation}\label{SC_KA anzac Sy} -\int_{C_{2}}^{\varphi}\exp\biggl(\frac{\imath\Omega_{\mu}}{\Omega_{e}}(\varphi'-\varphi)\biggr)(\Pi_{x}(\varphi')-\imath\Pi_{y}(\varphi'))d\varphi'\Biggr] \exp\biggl(-\imath\frac{(\omega-k_{z}v_{z})}{\Omega_{e}}\varphi\biggr). \end{equation}

Solutions for $\delta S_{x}$ and $\delta S_{y}$ (presented by equations (\ref{SC_KA anzac Sx}) and (\ref{SC_KA anzac Sy})) contain the following functions
$$\Pi_{x}(\varphi)=\frac{1}{\Omega_{e}}\exp\biggl(\imath\frac{(\omega-k_{z}v_{z})}{\Omega_{e}}\varphi\biggr)$$
\begin{equation}\label{SC_KA Pi x}  \times\biggl(\imath\mu_{e}(\textbf{k}\cdot\nabla_{\textbf{p}})f_{0}\delta
B_{x}+\frac{2\mu_{e}}{\hbar}S_{0z}\delta B_{y} +S_{0y}(\textbf{v}\cdot\delta \textbf{B})\frac{q_{e}}{c}\frac{v_{z}}{mv_{\perp}^{2}}
+q_{e}(\textbf{v}\cdot\delta \textbf{E})\partial_{\varepsilon}S_{0x}-q_{e}S_{0y}\frac{((\textbf{v}\times\delta \textbf{E})\textbf{e}_{z})}{mv_{\perp}^{2}}\biggr),\end{equation}
and
$$\Pi_{y}(\varphi)=\frac{1}{\Omega_{e}}\exp\biggl(\imath\frac{(\omega-k_{z}v_{z})}{\Omega_{e}}\varphi\biggr)$$
\begin{equation}\label{SC_KA Pi y}  \times\biggl(\imath\mu_{e}(\textbf{k}\cdot\nabla_{\textbf{p}})f_{0}\delta
B_{y}-\frac{2\mu_{e}}{\hbar}S_{0z}\delta B_{x} -S_{0x}(\textbf{v}\cdot\delta \textbf{B})\frac{q_{e}}{c}\frac{v_{z}}{mv_{\perp}^{2}} +q_{e}(\textbf{v}\cdot\delta \textbf{E})\partial_{\varepsilon}S_{0y}+q_{e}S_{0x}\frac{((\textbf{v}\times\delta \textbf{E})\textbf{e}_{z})}{mv_{\perp}^{2}}\biggr).\end{equation}

Constants $C_{0}$, $C_{1}$, $C_{2}$ and $C_{3}$ are chosen that distribution functions $\delta f$ and $\delta \textbf{S}$ are periodic functions of angle $\varphi$: $\delta f(\varphi+2\pi)=\delta f(\varphi)$ and $\delta \textbf{S}(\varphi+2\pi)=\delta \textbf{S}(\varphi)$.

\section{Appendix B: Explicit form of dielectric permeability tensor}

Tensor $\widetilde{\Pi}^{\alpha\beta}_{Cl}(\theta,s)$ has the following explicit form:
\begin{equation}\label{SC_KA Pi Cl} \widehat{\widetilde{\Pi}}_{Cl}(\theta,s)=\frac{3\omega_{Ls}^{2}}{2\omega}\left(\begin{array}{ccc}
\frac{1}{4}(\frac{\sin^{2}\theta}{\omega-k_{z}v_{Fs}\cos\theta-\Omega_{e}}+\frac{\sin^{2}\theta}{\omega-k_{z}v_{Fs}\cos\theta+\Omega_{e}}) & \imath \frac{1}{4}(\frac{\sin^{2}\theta}{\omega-k_{z}v_{Fs}\cos\theta-\Omega_{e}}-\frac{\sin^{2}\theta}{\omega-k_{z}v_{Fs}\cos\theta+\Omega_{e}}) & 0 \\
-\imath \frac{1}{4}(\frac{\sin^{2}\theta}{\omega-k_{z}v_{Fs}\cos\theta-\Omega_{e}}-\frac{\sin^{2}\theta}{\omega-k_{z}v_{Fs}\cos\theta+\Omega_{e}}) & \frac{1}{4}(\frac{\sin^{2}\theta}{\omega-k_{z}v_{Fs}\cos\theta-\Omega_{e}}+\frac{\sin^{2}\theta}{\omega-k_{z}v_{Fs}\cos\theta+\Omega_{e}}) & 0 \\
0 & 0 & \frac{\cos^{2}\theta}{\omega-k_{z}v_{Fs}\cos\theta}
\end{array}\right).\end{equation}\end{widetext}
Tensors $\varepsilon^{\alpha\beta}_{11}$ and $\varepsilon^{\alpha\beta}_{21}$ are calculated in Refs. \cite{Andreev PoP kinetics 17 a}, \cite{Andreev PoP kinetics 17 b} and have the following form: $\varepsilon^{\alpha\beta}_{11}=0$,
$$\varepsilon^{\alpha\beta}_{21,a}=-\frac{m^{2}}{\pi\hbar^{3}}\frac{\mu_{e}^{2}c^{2}}{2\omega^{2}}\times$$
\begin{equation}\label{SC_KA dielectric permeability tensor GF} \times\sum_{s=\uparrow, \downarrow}\int \sin\theta d\theta
\sum_{r=+,-} \frac{v_{Fs}^{2}k_{z}\cos\theta\kappa^{\alpha\beta}_{r}}{\omega-k_{z}v_{Fs}\cos\theta+r\Omega_{\mu}}
,\end{equation}
and
$$\varepsilon^{\alpha\beta}_{21,b}=\frac{m^{3}}{\pi\hbar^{3}}\frac{\mu_{e}^{2}c^{2}}{\hbar\omega^{2}}\times$$
\begin{equation}\label{SC_KA dielectric permeability tensor GF} \times\sum_{s=\uparrow, \downarrow}\int \sin\theta d\theta \sum_{r=+,-}\int_{0}^{v_{Fs}} \frac{r\kappa^{\alpha\beta}_{r} (-1)^{i_{s}} v^{2}dv}{\omega-k_{z}v \cos\theta+r\Omega_{\mu}},\end{equation}
where $\kappa^{\alpha\beta}_{-}=(K^{\alpha\beta}_{\parallel})^{*}$, $\kappa^{\alpha\beta}_{+}=K^{\alpha\beta}_{\parallel}$, $i_{\uparrow}=0$, $i_{\downarrow}=1$,
\begin{equation}\label{SC_KA } \hat{K}_{\parallel}=k_{z}^{2}\left(\begin{array}{ccc}
1 & -\imath & 0 \\
\imath & 1 & 0 \\
0 & 0 & 0 \\
\end{array}\right),\end{equation}
where $\omega_{Ls}^{2}=4\pi e^{2}n_{0s}/m$. $\widehat{\widetilde{\Pi}}_{Cl}(\theta,s)$ is similar to the traditional result for degenerate electrons presented in many textbooks (see for instance \cite{Rukhadze book 84}), but it also includes the spin separation effect.

Elements of the dielectric permeability tensor caused by the nontrivial part of $\delta f$ have the following structure
\begin{equation}\label{SC_KA } \varepsilon_{12}=\frac{4\pi}{\omega}\left(
                                                 \begin{array}{ccc}
\alpha_{+}-\alpha_{-} & -\imath(\alpha_{+}+\alpha_{-}) & 0 \\
\imath(\alpha_{+}+\alpha_{-}) & \alpha_{+}-\alpha_{-} & 0 \\
0 & 0 & 0 \\
\end{array}\right)
, \end{equation}
where
\begin{equation}\label{SC_KA } \alpha_{\pm}=\frac{1}{4}q_{e}\mu_{e}\frac{k_{z}^{2}c}{\omega}\int \frac{v_{z}v_{\perp}\partial_{p}\Sigma(p)}{\omega-k_{z}v_{z}\pm\Omega_{e}} \frac{d\textbf{p}}{v}, \end{equation}
and
\begin{equation}\label{SC_KA Sigma} \Sigma(p)=\frac{1}{(2\pi\hbar)^{3}}[\Theta(p_{F\uparrow}-p) -\Theta(p_{F\downarrow}-p)].\end{equation}

Elements of the dielectric permeability tensor caused by the nontrivial part of $\delta \textbf{S}$ have the following structure $\varepsilon_{22}=\varepsilon_{22,a}+\varepsilon_{22,b}+\varepsilon_{22,c}$:
\begin{equation}\label{SC_KA } \varepsilon_{22,a}= \left(
\begin{array}{ccc}
\kappa_{-}-\kappa_{+} & \imath(\kappa_{+}+\kappa_{-}) & 0 \\
-\imath(\kappa_{+}+\kappa_{-}) & \kappa_{-}-\kappa_{+} & 0 \\
0 & 0 & 0 \\
\end{array}\right),\end{equation}

\begin{equation}\label{SC_KA } \varepsilon_{22,b}= \left(
\begin{array}{ccc}
\chi_{+}-\chi_{-} & -\imath(\chi_{+}+\chi_{-}) & 0 \\
\imath(\chi_{+}+\chi_{-}) & \chi_{+}-\chi_{-} & 0 \\
0 & 0 & 0 \\
\end{array}\right),\end{equation}
and
\begin{equation}\label{SC_KA } \varepsilon_{22,c}=\frac{4\pi}{\omega}\left(
\begin{array}{ccc}
\beta_{+}-\beta_{-} & -\imath(\beta_{+}+\beta_{-}) & 0 \\
\imath(\beta_{+}+\beta_{-}) & \beta_{+}-\beta_{-} & 0 \\
0 & 0 & 0 \\
\end{array}\right),\end{equation}
where
\begin{equation}\label{SC_KA beta interm} \beta_{\pm}=\frac{q_{e}\mu_{e}}{4m}\frac{k_{z}^{2}c}{\omega}\int \frac{v_{z}}{v_{\perp}}\frac{\Sigma(p)}{\omega-k_{z}v_{z}\pm\Omega_{\mu}}d\textbf{p}, \end{equation}
\begin{equation}\label{SC_KA kappa interm} \kappa_{\pm}=\frac{q_{e}\mu_{e}}{(2\pi\hbar)^{3}}\frac{k_{z}c}{\omega}\sum_{s}\int\frac{2\pi^{2}(-1)^{i_{s}}p_{Fs}^{2}\sin^{2}\theta d\theta}{\omega-k_{z}v_{Fs}\cos\theta\pm\Omega_{\mu}},\end{equation}
and
\begin{equation}\label{SC_KA chi interm}\chi_{\pm}=\frac{k_{z}c}{m\omega}\frac{\pi q_{e}\mu_{e}}{(2\pi\hbar)^{3}}\sum_{s}(-1)^{i_{s}}\int \frac{d\textbf{p}}{v_{\perp}}\frac{\Theta(p_{Fs}-p)}{\omega-k_{z}v_{z}\pm\Omega_{\mu}}. \end{equation}

Further integration in the dielectric permeability tensor gives the following result:
\begin{equation}\label{SC_KA } \varepsilon^{\alpha\beta}_{10}=-\sum_{s=\uparrow, \downarrow} \widetilde{\Pi}^{\alpha\beta}_{Cl}(s), \end{equation}
with
$$\widehat{\widetilde{\Pi}}_{Cl}(s)=\frac{3\omega_{Ls}^{2}}{2\omega}\times$$
\begin{equation}\label{SC_KA Pi Cl} \times\left(\begin{array}{ccc}
\frac{1}{4}[G_{-}+G_{+}] & \imath \frac{1}{4}[G_{-}-G_{+}] & 0 \\
-\imath \frac{1}{4}[G_{-}-G_{+}] & \frac{1}{4}[G_{-}+G_{+}] & 0 \\
0 & 0 & G_{zz}
\end{array}\right),\end{equation}
where
$$G_{\pm}=G(\omega\pm\Omega_{e})=
\frac{1}{k_{z}v_{Fs}}\biggl[\frac{2(\omega\pm\Omega_{e})}{k_{z}v_{Fs}}$$
\begin{equation}\label{SC_KA }
+\biggl(1-\frac{(\omega\pm\Omega_{e})^{2}}{(k_{z}v_{Fs})^{2}}\biggr)\ln\biggl(\frac{\omega+k_{z}v_{Fs}\pm\Omega_{e}}{\omega-k_{z}v_{Fs}\pm\Omega_{e}}\biggr)\biggr], \end{equation}
and
\begin{equation}\label{SC_KA } G_{zz}=\frac{\omega}{(k_{z}v_{Fs})^{2}}\biggl[\frac{\omega}{k_{z}v_{Fs}}\ln\biggl(\frac{\omega+k_{z}v_{Fs}}{\omega-k_{z}v_{Fs}}\biggr)-2\biggr]. \end{equation}

Tensors $\varepsilon_{11}^{\alpha\beta}$ and $\varepsilon_{21}^{\alpha\beta}$ are caused by contribution of $f_{0}$ and $S_{0z}$ in the spin evolution. They are derived in \cite{Andreev PoP kinetics 17 a} and \cite{Andreev PoP kinetics 17 b}. Their structure is described in the following form:
\begin{equation}\label{SC_KA } \widehat{\varepsilon}_{21,a}=k_{z}^{2}\left(\begin{array}{ccc}
\gamma_{+}+\gamma_{-} & -\imath(\gamma_{+}-\gamma_{-}) & 0 \\
\imath(\gamma_{+}-\gamma_{-}) & \gamma_{+}+\gamma_{-} & 0 \\
0 & 0 & 0 \\
\end{array}\right), \end{equation}
and
\begin{equation}\label{SC_KA } \widehat{\varepsilon}_{21,b}=k_{z}^{2}\left(\begin{array}{ccc}
\delta_{+}-\delta_{-} & -\imath(\delta_{+}+\delta_{-}) & 0 \\
\imath(\delta_{+}+\delta_{-}) & \delta_{+}-\delta_{-} & 0 \\
0 & 0 & 0 \\
\end{array}\right) .\end{equation}

Next, present the explicit forms of elements $\gamma_{\pm}$ and $\delta_{\pm}$
$$\gamma_{\pm}=\sum_{s=\uparrow, \downarrow} \frac{m^{2}v_{Fs}}{\pi\hbar^{3}}\frac{\mu_{e}^{2}c^{2}}{2\omega^{2}}\times$$
\begin{equation}\label{SC_KA } \times
\Biggl(2-\frac{\omega \pm\Omega_{\mu}}{k_{z}v_{Fs}} \ln\biggl(\frac{\omega+k_{z}v_{Fs}\pm\Omega_{\mu}}{\omega-k_{z}v_{Fs}\pm\Omega_{\mu}}\biggr)\Biggr), \end{equation}
and
$$\delta_{\pm}=\sum_{s=\uparrow, \downarrow}\frac{m^{3}}{\pi\hbar^{3}}\frac{\mu_{e}^{2}c^{2}}{\hbar\omega^{2}} \frac{(-1)^{i_{s}}}{k_{z}^{2}}\Biggl(v_{Fs}(\omega\pm\Omega_{\mu})$$
\begin{equation}\label{SC_KA }  -\frac{(\omega\pm\Omega_{\mu})^{2}-(k_{z}v_{Fs})^{2}}{2k_{z}} \ln\biggl(\frac{\omega+k_{z}v_{Fs}\pm\Omega_{\mu}}{\omega-k_{z}v_{Fs}\pm\Omega_{\mu}}\biggr)\Biggr). \end{equation}

The integral in $\alpha_{\pm}$ contains the Dirac delta function of the momentum module. So, it can be easily presented as an integral over angle $\theta$:
$$\alpha_{\pm}=-\frac{\pi q_{e}\mu_{e}}{2(2\pi\hbar)^{3}}\frac{k_{z}^{2}c}{\omega}\sum_{s}(-1)^{i_{s}}\times$$
\begin{equation}\label{SC_KA } \times m^{2}v_{Fs}^{3}\int\frac{\sin^{2}\theta\cos\theta d\theta}{\omega-k_{z}v_{Fs}\cos\theta\pm\Omega_{e}}.\end{equation}
After taking the last integral, the explicit form of $\alpha_{\pm}$ is found:
$$\alpha_{\pm}=-\frac{q_{e}\mu_{e}}{16\pi\hbar^{3}}\frac{k_{z}c}{\omega}\sum_{s}(-1)^{i_{s}} m^{2}v_{Fs}^{2}  \biggl[-\frac{1}{2}+\biggl(\frac{\omega\pm\Omega_{e}}{k_{z}v_{Fs}}\biggr)^{2}$$
\begin{equation}\label{SC_KA }
-\frac{\omega\pm\Omega_{e}}{k_{z}v_{Fs}}\biggl(1-\frac{\omega\pm\Omega_{e}}{k_{z}v_{Fs}}\biggr) \sqrt{\frac{\omega\pm\Omega_{e}+k_{z}v_{Fs}}{\omega\pm\Omega_{e}-k_{z}v_{Fs}}} \biggr]. \end{equation}

Using the explicit form of $\Sigma(p)$ represent $\beta_{\pm}$ in the following form:
$$\beta_{\pm}=\frac{k_{z}c}{\omega}\frac{q_{e}\mu_{e}}{16\pi^{2}\hbar^{3}}\sum_{s}(-1)^{i_{s}}\times$$
\begin{equation}\label{SC_KA }\times \int_{0}^{\pi} d\theta\int_{0}^{p_{Fs}} pdp\frac{\cos\theta}{\frac{\omega\pm\Omega_{\mu}}{k_{z}v}-\cos\theta}.\end{equation}
Taking integral over the angle $\theta$ find the following:
$$\beta_{\pm}=\frac{k_{z}c}{\omega}\frac{q_{e}\mu_{e}}{16\pi\hbar^{3}}\sum_{s}(-1)^{i_{s}}\int_{0}^{p_{Fs}} pdp\biggl(-1$$
\begin{equation}\label{SC_KA }+ \biggl(\frac{\omega\pm\Omega_{\mu}}{\omega\pm\Omega_{\mu}+k_{z}v}\biggr) \sqrt{\frac{\omega\pm\Omega_{\mu}+k_{z}v}{\omega\pm\Omega_{\mu}-k_{z}v}}\biggr).\end{equation}
Finally, taking integral over the momentum module find the explicit form of function $\beta_{\pm}$:
$$\beta_{\pm}=\frac{q_{e}\mu_{e}}{16\pi\hbar^{3}}\frac{k_{z}c}{\omega}\sum_{s}(-1)^{i_{s}}\Biggl(-\frac{1}{2}p_{Fs}^{2} $$
\begin{equation}\label{SC_KA beta final 1} +\frac{m^{2}}{k_{z}^{2}}(\omega\pm\Omega_{\mu})\biggl( (\omega\pm\Omega_{\mu})-\sqrt{(\omega\pm\Omega_{\mu})^{2}-k_{z}^{2}v_{Fs}^{2}}\biggr)\Biggr) \end{equation}
for $\omega\pm\Omega_{\mu}>0$ and $\omega+\Omega_{\mu}<-k_{z}v_{Fs}$,
or
$$\beta_{+}=\frac{q_{e}\mu_{e}}{16\pi\hbar^{3}}\frac{k_{z}c}{\omega}\sum_{s}(-1)^{i_{s}}\Biggl(-\frac{1}{2}p_{Fs}^{2} $$
\begin{equation}\label{SC_KA beta final 1}
+\frac{m^{2}}{k_{z}^{2}}(\omega+\Omega_{\mu})\biggl( (\omega+\Omega_{\mu})+\sqrt{(\omega+\Omega_{\mu})^{2}-k_{z}^{2}v_{Fs}^{2}}\biggr)\Biggr) \end{equation}
for $-k_{z}v_{Fs}<\omega+\Omega_{\mu}<0$.

The final form of the functions $\kappa_{\pm}$ can be found after integration over the angle in equation (\ref{SC_KA kappa interm}).
It appears as follows 
$$\kappa_{\pm}=q_{e}\mu_{e}\frac{k_{z}c}{\omega}\frac{1}{k_{z}}\frac{m^{2}}{4\hbar^{3}}\sum_{s}(-1)^{i_{s}}v_{Fs}\times$$
\begin{equation}\label{SC_KA }  \times\biggl[\frac{\omega\pm\Omega_{\mu}}{k_{z}v_{Fs}}+\biggl(1-\frac{\omega\pm\Omega_{\mu}}{k_{z}v_{Fs}}\biggr) \sqrt{\frac{\omega\pm\Omega_{\mu}+k_{z}v_{Fs}}{\omega\pm\Omega_{\mu}-k_{z}v_{Fs}}}\biggr]. \end{equation}  

To find the final form of functions $\chi_{\pm}$ take integrals over angles $\varphi$ and $\theta$ and then take integral over the velocity module and obtain the following results: 
$$\chi_{\pm}=q_{e}\mu_{e}\frac{c}{\omega}\frac{m^{2}}{4\hbar^{3}}\sum_{s}(-1)^{i_{s}}\times$$ $$\times\int_{0}^{v_{Fs}}dv\frac{1}{1+\frac{\omega\pm\Omega_{\mu}}{k_{z}v}} \sqrt{\frac{\omega\pm\Omega_{\mu}+k_{z}v}{\omega\pm\Omega_{\mu}-k_{z}v}}$$
$$=q_{e}\mu_{e}\frac{k_{z}c}{\omega}\frac{1}{k_{z}^{2}}\frac{m^{2}}{2\hbar^{3}}\sum_{s}(-1)^{i_{s}} \biggl(\omega\pm\Omega_{\mu}$$
\begin{equation}\label{SC_KA } -\sqrt{(\omega\pm\Omega_{\mu})^{2}-k_{z}^{2}v_{Fs}^{2}}\biggr). \end{equation}

The dispersion equation appears in the following form at the application of the found structure of the dielectric permeability tensor:
\begin{equation}\label{SC_KA dielectric permeability tensor as det}det\left(
\begin{array}{ccc}
\varepsilon_{xx}-\frac{k_{z}^{2}c^{2}}{\omega^{2}} & \varepsilon_{xy} & 0 \\
    \varepsilon_{yz} & \varepsilon_{yy}-\frac{k_{z}^{2}c^{2}}{\omega^{2}} & 0 \\
    0 & 0 & \varepsilon_{zz} \\
\end{array}\right)=0,\end{equation}
with $\varepsilon_{xx}=\varepsilon_{yy}\equiv\epsilon$, and $\varepsilon_{yx}^{*}=\varepsilon_{xy}=\imath\Xi$, where
$\epsilon=k_{z}^{2}(\gamma_{+}+\gamma_{-}+\delta_{+}-\delta_{-})+\frac{4\pi}{\omega}(\alpha_{+}-\alpha_{-}+\beta_{+}-\beta_{-})
+\kappa_{-}-\kappa_{+}-\chi_{-}+\chi_{+}
-\sum_{s}\frac{3}{8}\frac{\omega_{Ls}^{2}}{\omega}(G_{+}+G_{-})$,
$\Xi=-k_{z}^{2}(\gamma_{+}-\gamma_{-}+\delta_{+}+\delta_{-}) -\frac{4\pi}{\omega}(\alpha_{+}+\alpha_{-}+\beta_{+}+\beta_{-})
+\kappa_{-}+\kappa_{+}-\chi_{-}-\chi_{+}
-\sum_{s}\frac{3}{8}\frac{\omega_{Ls}^{2}}{\omega}(G_{-}-G_{+})$.
All functions
$\alpha_{\pm}$, $\beta_{\pm}$, $\gamma_{\pm}$, $\delta_{\pm}$, $\kappa_{\pm}$, $\chi_{\pm}$, $G_{\pm}$
are described above, these functions are related to the following elements of the dielectric permeability tensor
$\varepsilon_{12}$, $\varepsilon_{22,c}$, $\varepsilon_{21,a}$, $\varepsilon_{21,b}$, $\varepsilon_{22,a}$, $\varepsilon_{22,b}$, $\varepsilon_{10}$ correspondingly,
where $\varepsilon_{21}=\varepsilon_{21,a}+\varepsilon_{21,b}$, $\varepsilon_{22}=\varepsilon_{22,a}+\varepsilon_{22,b}+\varepsilon_{22,c}$ and $\varepsilon_{11}=0$ for the waves propagating parallel to the external field.
Tensor $\varepsilon_{21,a}$ comes from $\mu_{e}\nabla_{\textbf{r}}^{\alpha}\delta B^{\alpha}\cdot\nabla_{\textbf{p}}^{\beta}f_{0}$ in equation (\ref{SC_KA kinetic equation gen for spin classic limit with E and B}).
Tensor $\varepsilon_{21,b}$ comes from torque-like term in equation (\ref{SC_KA kinetic equation gen for spin classic limit with E and B}): $-(2\mu_{e}/\hbar)S_{0z}(\textbf{r},\textbf{p},t)\textbf{e}_{z}\times\delta\textbf{B}(\textbf{r},t)$.
Tensors $\varepsilon_{22,a}$ and $\varepsilon_{22,b}$ appear from $q_{e}(\delta \textbf{E}\cdot\nabla_{\textbf{p}})S_{0x}$ and $q_{e}(\delta \textbf{E}\nabla_{\textbf{p}})S_{0y}$ in equation (\ref{SC_KA kinetic equation gen for spin classic limit with E and B}).
Tensor $\varepsilon_{22,a}$ comes from $q_{e}(\delta \textbf{E}\cdot \textbf{v})\partial_{\varepsilon}S_{0x}$ and $q_{e}(\delta \textbf{E}\cdot \textbf{v})\partial_{\varepsilon}S_{0y}$.
Tensor $\varepsilon_{22,b}$ comes from $(\textbf{e}_{z}\times\textbf{S}_{0})((\textbf{v} \times\delta \textbf{E})\cdot\textbf{e}_{z})/mv_{\perp}^{2}$
Tensor $\varepsilon_{22,c}$ appears from $(q_{e}/c)((\textbf{v}\times\delta \textbf{B})\nabla_{\textbf{p}})\textbf{S}_{0}$.
All $\pm$ and $\mp$ presented above describe parts of terms containing $\omega+\Omega_{a}$ or $\omega-\Omega_{a}$, where $a=e, \mu$. The dielectric permeability tensor (\ref{SC_KA dielectric permeability tensor as det}) contains superposition of these terms.

The dispersion equation splits on three equations, one equation is for the longitudinal waves $\varepsilon_{zz}=0$
which is discussed in several papers \cite{Andreev PoP kinetics 17 b}, \cite{Andreev PoP 16 sep kin}
and two equations are for the transverse waves
\begin{equation}\label{SC_KA tr waves general form} \frac{k_{z}^{2}c^{2}}{\omega^{2}}=\epsilon\pm\Xi,\end{equation}
where
$$\epsilon+\Xi=2k_{z}^{2}(\gamma_{-}-\delta_{-})-\frac{8\pi}{\omega}(\alpha_{-}+\beta_{-})$$
\begin{equation}\label{SC_KA} +2\kappa_{-}-2\chi_{-} -\sum_{s}\frac{3}{4}\frac{\omega_{Ls}^{2}}{\omega}G_{-}, \end{equation}
and
$$\epsilon-\Xi=2k_{z}^{2}(\gamma_{+}+\delta_{+})+\frac{8\pi}{\omega}(\alpha_{+}+\beta_{+})$$
\begin{equation}\label{SC_KA} -2\kappa_{+}+2\chi_{+} -\sum_{s}\frac{3}{4}\frac{\omega_{Ls}^{2}}{\omega}G_{+}. \end{equation}
Coefficient $\pm$ in equation (\ref{SC_KA tr waves general form}) appears as solution of a quadratic equation. Hence, this coefficient is independent from all $\pm$ and $\mp$ presented above. In equation (\ref{SC_KA tr waves general form}) $\pm$ corresponds to transverse waves with different circular polarizations $\delta E_{x}=\pm\imath\delta E_{y}$. Here and below $\pm$ and $\mp$ are produced by $\pm$ in equation (\ref{SC_KA tr waves general form}).

Presented above leads to the following structure of function $\Sigma_{\mp}$ introduced in equation (\ref{SC_KA tr waves explicit form})
\begin{equation}\label{SC_KA} \Sigma_{\mp}=2k_{z}^{2}(\gamma_{\mp}\mp\delta_{\mp})\mp\frac{8\pi}{\omega}(\alpha_{\mp}+\beta_{\mp}) \pm2\kappa_{\mp}\mp2\chi_{\mp}. \end{equation}

\section{Appendix C: Approximate form of functions $\alpha_{\pm}$, $\beta_{\pm}$, $\gamma_{\pm}$, $\delta_{\pm}$}

Consider the small frequency and small wave vector limit, where $k_{z}v_{Fs}/\mid\omega\pm\Omega_{e}\mid\ll1$.

Consider approximate forms of functions $\alpha_{\pm}$, $\beta_{\pm}$, $\gamma_{\pm}$, $\delta_{\pm}$ which are elements of dispersion equation (\ref{SC_KA tr waves explicit form}):
$$\widetilde{\varepsilon}_{10}=1 -\sum_{s=\uparrow, \downarrow}\frac{\omega_{Ls}^{2}}{\omega k_{z}v_{Fs}}\biggl[\frac{k_{z}v_{Fs}}{\omega\pm\mid\Omega_{e}\mid}
+\frac{1}{5}\biggl(\frac{k_{z}v_{Fs}}{\omega\pm\mid\Omega_{e}\mid}\biggr)^{3}\biggr]$$
\begin{equation}\label{SC_KA}\approx 1-\frac{\omega_{Ls}^{2}}{\omega}\frac{1}{\omega\pm\mid\Omega_{e}\mid},\end{equation}

\begin{equation}\label{SC_KA} \widetilde{\varepsilon}_{21,a}=2k_{z}^{2}\gamma_{\mp}= -\frac{\omega_{Ls}^{2}}{\omega^{2}}\frac{\hbar^{2}k_{z}^{4}}{m^{2}}\frac{1}{(\omega\pm\mid\Omega_{\mu}\mid)^{2}},\end{equation}

$$\widetilde{\varepsilon}_{21,b}=\mp 2k_{z}^{2}\delta_{\mp}$$
$$=\sum_{s=\uparrow, \downarrow}(-1)^{i_{s}} \frac{\omega_{Ls}^{2}}{\omega^{2}}\frac{\hbar k_{z}}{2mv_{Fs}} \biggl[\frac{k_{z}v_{Fs}}{\omega\pm\mid\Omega_{\mu}\mid}
+\frac{1}{5}\biggl(\frac{k_{z}v_{Fs}}{\omega\pm\mid\Omega_{\mu}\mid}\biggr)^{3}\biggr]$$
\begin{equation}\label{SC_KA}\approx \frac{\omega_{Lu}^{2}-\omega_{Ld}^{2}}{\omega^{2}}\frac{\hbar k_{z}^{2}}{2m} \frac{1}{\omega\pm\mid\Omega_{\mu}\mid},\end{equation}

$$\widetilde{\varepsilon}_{12}=\mp\frac{8\pi}{\omega}\alpha_{\mp}$$
\begin{equation}\label{SC_KA wt varepsilon 12} =\mp\frac{3\pi(6\pi^{2})^{\frac{1}{3}}}{16} \frac{\omega_{Le}^{2}}{\omega^{2}}\frac{\hbar^{2}k_{z}^{3}}{4m^{2}} \frac{n_{0u}^{\frac{4}{3}}-n_{0d}^{\frac{4}{3}}}{n_{0e}(\omega\pm\mid\Omega_{e}\mid)^{2}} , \end{equation} 

$$\widetilde{\varepsilon}_{22,c}= \mp\frac{8\pi}{\omega}\beta_{\mp}$$
\begin{equation}\label{SC_KA wt varepsilon 22 c}
=\mp\frac{3\pi(6\pi^{2})^{\frac{1}{3}}}{16} \frac{\omega_{Le}^{2}}{\omega^{2}}\frac{\hbar^{2}k_{z}^{3}}{4m^{2}} \frac{n_{0u}^{\frac{4}{3}}-n_{0d}^{\frac{4}{3}}}{n_{0e}(\omega\pm\mid\Omega_{\mu}\mid)^{2}}   ,\end{equation}
at $\omega\pm\Omega_{\mu}>0$ and $\omega+\Omega_{\mu}<0$,
$\widetilde{\varepsilon}_{12}$ and $\widetilde{\varepsilon}_{22,c}$ presented by equations (\ref{SC_KA wt varepsilon 12}) and (\ref{SC_KA wt varepsilon 22 c}) have same form, but differs by the different cyclotron frequencies entering their expressions,

$$\widetilde{\varepsilon}_{22,a}=\pm 2\kappa_{\mp}= \pm 2q_{e}\mu_{e} \frac{k_{z}c}{\omega}\frac{1}{k_{z}}\frac{m^{2}}{8\hbar^{3}}\sum_{s}(-1)^{i_{s}}v_{Fs}\times$$
$$\times \biggl[\frac{k_{z}v_{Fs}}{\omega\pm\mid\Omega_{\mu}\mid} +\frac{1}{4}\frac{k_{z}^{3}v_{Fs}^{3}}{(\omega\pm\mid\Omega_{\mu}\mid)^{3}}\biggr]$$
\begin{equation}\label{SC_KA} \approx \pm\frac{(6\pi^{2})^{\frac{2}{3}}}{32\pi} \frac{\omega_{Le}^{2}}{\omega (\omega\pm\mid\Omega_{\mu}\mid)} \frac{n_{0u}^{\frac{2}{3}}-n_{0d}^{\frac{2}{3}}}{n_{0e}}k_{z}, \end{equation}

If $\omega\pm\mid\Omega_{\mu}\mid >0$, the functions $\chi_{\mp}$ give the following assumptions:
$$\widetilde{\varepsilon}_{22,b}=\mp 2\chi_{\mp}=  \mp 2 q_{e}\mu_{e} \frac{k_{z}c}{\omega}\frac{1}{k_{z}}\frac{m^{2}}{4\hbar^{3}}\sum_{s}(-1)^{i_{s}}v_{Fs}\times$$
$$\times \frac{k_{z}v_{Fs}}{\omega\pm\mid\Omega_{\mu}\mid}\biggl(1+\frac{1}{4}\frac{k_{z}^{2}v_{Fs}^{2}}{(\omega\pm\mid\Omega_{\mu}\mid)^{2}}\biggr)$$
\begin{equation}\label{SC_KA} \approx \mp\frac{(6\pi^{2})^{\frac{2}{3}}}{16\pi} \frac{\omega_{Le}^{2}}{\omega (\omega\pm\mid\Omega_{\mu}\mid)} \frac{n_{0u}^{\frac{2}{3}}-n_{0d}^{\frac{2}{3}}}{n_{0e}}k_{z}\approx \mp4\kappa_{\mp}. \end{equation}
If $\omega-\mid\Omega_{\mu}\mid<0$, the function $\chi_{+}$ give the following assumption:
$$\widetilde{\varepsilon}_{22,b}=2\chi_{+}=
\frac{k_{z}c}{\omega}\frac{q_{e}\mu_{e}}{k_{z}^{2}}\frac{m^{2}}{\hbar^{3}}\times$$
\begin{equation}\label{SC_KA} \times\sum_{s}(-1)^{i_{s}} \biggl(2(\omega-\mid\Omega_{\mu}\mid)-\frac{1}{2}\frac{k_{z}^{2}v_{Fs}^{2}}{\omega-\mid\Omega_{\mu}\mid}\biggr)\approx0.\end{equation}

Functions $\alpha_{\pm}$, $\beta_{\pm}$ and $\gamma_{\pm}$ gives no contribution in equation (\ref{SC_KA tr waves explicit form expanded  reduced 2}). Functions $\delta_{\pm}$ give the third term on the right-hand side. Functions $\kappa_{\pm}$ and $\chi_{\pm}$ lead to the last term. Earlier result can be found for instance in  Ref. \cite{Misra JPP 10}. The fourth term is an extra term in compare with earlier papers.

\section{Appendix D: Dimensionless form of the dispersion equation for transverse waves}

Dimensionless variables are used for numerical analysis of the obtained results
$\xi=\omega/\omega_{Le}$, $\kappa=k_{z}c/\omega_{Le}$, $f=\mid\Omega_{e}\mid/\omega_{Le}$, $g=1.00116$.
Equation (\ref{SC_KA tr waves explicit form expanded  reduced 2}) is presented in the described dimensionless variables:
$$\xi^{2}-\kappa^{2}-\frac{\xi}{\xi\pm f}-\eta \Upsilon\kappa^{2}\frac{1}{\xi\pm g f}$$
\begin{equation}\label{SC_KA tr waves explicit form expanded  reduced 3} \pm\frac{(3\pi^{2})^{\frac{2}{3}}}{32\pi}((1+\eta)^{\frac{2}{3}}-(1-\eta)^{\frac{2}{3}})\frac{\kappa}{R}
\frac{\xi}{\xi\pm g f}=0 \end{equation}
for $\xi\pm f>0$, where $R=\sqrt[3]{n_{0}}c/\omega_{Le}$, $\Upsilon=\hbar\omega_{Le}/mc^{2}$, for $n_{0e}=10^{27}$ cm$^{-3}$ find $R\approx16,7$, $\Upsilon\approx2.4\times10^{-3}$. Signs in equation (\ref{SC_KA tr waves explicit form expanded  reduced 3}) corresponds to the circular polarization of the plane wave $\delta E_{x}=\pm\imath\delta E_{y}$.
If $\xi-f<0$ equation (\ref{SC_KA tr waves explicit form expanded  reduced 3}) changes to
$$\xi^{2}-\kappa^{2}-\frac{\xi}{\xi- f}-\eta \Upsilon\kappa^{2}\frac{1}{\xi- g f}$$
\begin{equation}\label{SC_KA tr waves explicit form expanded  reduced 4}
+\frac{(3\pi^{2})^{\frac{2}{3}}}{32\pi}((1+\eta)^{\frac{2}{3}}-(1-\eta)^{\frac{2}{3}})
\frac{\kappa}{R}\frac{\xi}{\xi- g f}=0. \end{equation}

\end{document}